\documentclass[%
 reprint,
superscriptaddress,
 amsmath,amssymb,
 aps,
prb,
longbibliography
]{revtex4-2}

\usepackage[T1]{fontenc} 
\usepackage{graphicx}
\usepackage{amsmath}
\usepackage{amssymb}
\usepackage{textgreek}
\usepackage{tabularx}

\newcommand{\Deff}{D_\mathrm{eff}}
\newcommand{\Ms}{M_\mathrm{S}}
\newcommand{\Ku}{K_\mathrm{U}}
\newcommand{\Aex}{A}
\newcommand{\Hueff}{H_\mathrm{Ueff}}
\newcommand{\fs}{f_\mathrm{S}}
\newcommand{\fas}{f_\mathrm{aS}}

\begin{document}

\title{Spin wave frequency hysteresis in Ir/Co/Pt multilayers with Dzyaloshinskii–Moriya interaction}
\author{Ryszard Gieniusz}\email{gieniusz@uwb.edu.pl}
\affiliation{Faculty of Physics, University of Bialystok, Bialystok, Poland}

\author{Pawel Gruszecki}
\affiliation{Faculty of Physics, Adam Mickiewicz University, Poznan, Poland}
\author{Jan Kisielewski}\email{jankis@uwb.edu.pl}
\affiliation{Faculty of Physics, University of Bialystok, Bialystok, Poland}
\author{Anuj Kumar Dhiman}
\affiliation{Faculty of Physics, Adam Mickiewicz University, Poznan, Poland}
\author{Michal Matczak}
\affiliation{Institute of Molecular Physics, Polish Academy of Sciences, Poznan, Poland}
\author{Zbigniew Kurant}
\affiliation{Faculty of Physics, University of Bialystok, Bialystok, Poland}
\author{Iosif Sveklo}
\affiliation{Faculty of Physics, University of Bialystok, Bialystok, Poland}
\author{Urszula Guzowska}
\affiliation{Faculty of Physics, University of Bialystok, Bialystok, Poland}
\author{Maria Tekielak}
\affiliation{Faculty of Physics, University of Bialystok, Bialystok, Poland}
\author{Maciej Krawczyk}
\affiliation{Faculty of Physics, Adam Mickiewicz University, Poznan, Poland}
\author{Feliks Stobiecki}
\affiliation{Institute of Molecular Physics, Polish Academy of Sciences, Poznan, Poland}
\author{Andrzej Maziewski}
\affiliation{Faculty of Physics, University of Bialystok, Bialystok, Poland}

\begin{abstract}

Results of extensive combined experimental and theoretical investigations on static and dynamic properties of Ir/Co/Pt multilayer with low uniaxial anisotropy and asymmetric Ir/Co and Co/Pt interfaces responsible for large interfacial Dzyaloshinskii–Moriya interaction (IDMI) are presented. Within longitudinal magneto-optical Kerr effect-based microscopy and magnetic force microscopy studies a complex magnetic configuration was detected: large in-plane magnetized domains of several dozen micrometers size were modulated by a weak stripe domain pattern with periods of about 100 nm. Using Brillouin Light Scattering spectrometry, the hysteresis of the Stokes and anti-Stokes peaks frequencies was observed as a function of the magnetic field. This hysteretic behavior associated with IDMI-induced asymmetry of spin waves dispersion is correlated with switching of the large macro-domains. Using micromagnetic simulations we determine field-dependent magnetization distributions and dispersion relations, proposing an explanation of the observed behavior. The investigated nanostructure can be used as non-volatile spin waves velocity switcher. 


\end{abstract}

\maketitle

\section{Introduction}

Magnetic ultrathin films and multilayers are exciting systems with properties widely driven by different methods \cite{Johnson1996RPP, Saenz2020NT, Hellman2017RMP}. For instance, while decreasing thickness ($d$) of an ultrathin magnetic film, N\'eel surface anisotropy enables an increase of the uniaxial magnetic anisotropy, resulting in spin reorientation transition (SRT) from the in-plane to out-of-plane magnetization state. Two commonly used parameters for SRT description are: (i) the quality factor $Q = K_U/(\frac{1}{2}\mu_0\Ms^2)$, i.e. the ratio of the magnetic uniaxial anisotropy and demagnetization energies, where $\Ku$ and $\Ms$ are the uniaxial anisotropy constant and the magnetization saturation, respectively, and (ii) the effective uniaxial anisotropy field $\Hueff= \Ms (Q-1)$. In ultrathin films the magnetization distribution is homogeneous across the film thickness, and the SRT occurs at $Q\approx1$ (when $\Hueff$ changes sign) \cite{kisielewski02prl, maziewski14pssa}. Interestingly, while increasing $d$, the second SRT called as ``high-thickness SRT'' can occur (e.g., for Co it corresponds to few tens of nm). In this case, however, the resulting internal domain structure has small lateral periodicity with narrow domain walls of flux closure magnetization distribution, like in-plane-oriented vortices \cite{hubert98,kisielewski07jmmm}. 
Magnetization dynamics of resonant modes in such kind of weak stripe domains was already studied for several decades, both theoretially \cite{vukadinovic2000prl} and experimentally \cite{vukadinovic2000prl,talbi2010jpcs,camara17jpcm}, and oscillation modes in either domains or domain walls were identified. Particularly, domain walls attract a great interest, as a proper media for the propagation of spin waves (SWs), being considered as a ultra-narrow waveguides allowing easily to direct SWs in desired direction \cite{yu21pr, wagner16nat, xing16npg, xing17pra, lan15prx, garcia15prl}. Moreover, since both the amplitude and phase of the SWs change while passing through a single domain wall \cite{hertel04prl, borys2016aem, han19sci}, periodically aligned stripe domain patterns can be used as reconfigurable magnonic crystals with lattice constant unachievable by other techniques \cite{borys2016aem, banerjee17prb, gruszecki19ssp, tacchi14prb, camara17jpcm, grassi21, szulc2022}.
 
A lot of attention has been paid recently on the interfacial Dzyaloshinskii-Moriya interaction (IDMI) \cite{Dzi58,Mor60,Fer23,bode07nat,fert17nat}. IDMI is an antisymmetric exchange interaction between neighboring magnetic atoms that occurs at the interface between magnetic and covering nonmagnetic layers. The value and sign of IDMI depends on used covering material, so IDMI of the system can be substantially increased by selecting an appropriate covering pair, such as Pt and Ir layers \cite{fert17nat,ishikuro19prb}. IDMI favors orthogonal alignment of the neighboring spins, resulting in chiral magnetic structures, such as N\'eel type domain wall \cite{thiaville12epl, shen19jmmm}, spin spirals \cite{finco17prl, han16nl}, and skyrmions \cite{qiu20jmmm, you15cap, wiesendanger16nat, mruczkiewicz18prb}. Possible magnetization states have been studied by simulations in ultrathin films, considering wide range of both IDMI and magnetic anisotropy constant \cite{kisielewski19njp}. IDMI also influences the thickness-induced SRT, which can then occur for $Q$ well below 1. On the other hand, skyrmions are mainly studied for layers with $Q > 1$.

The presence of IDMI in materials with out-of-plane magnetization ($Q > 1$) results in asymmetric domain wall propagation \cite{je13prb, hrabec14prb, kuswik18prb} and changes the shape of hysteresis loops of laterally asymmetric microstructures \cite{kidanemariam19poly}. These effects are used to determine the magnitude of IDMI. Apart from having a significant impact on static magnetization textures, it also notably affects the SWs dynamics. Even in uniformly magnetized thin layers it introduces nonreciprocity in SWs dispersion, which is particularly visible for so-called Damon-Eschbach (DE) configuration \cite{koerner15prb, cho2015nat, moon13prb}, i.e., when SWs propagate perpendicularly to the in-plane applied magnetic field ($H$). In other words, it makes the SWs dispersion relation asymmetric – the SWs of a given wavelength may exhibit substantially different frequency for opposite propagation directions. This nonreciprocity is directly measurable by Brillouin light scattering (BLS) spectroscopy, as a difference between frequencies of Stokes ($\fs$) and anti-Stokes ({$\fas$) peaks. BLS spectroscopy does not requires special sample preparation, it can be used for samples with both in-plane and out-of-plane magnetization, so it became the most direct and straightforward method for IDMI determination \cite{kuepferling2023rmp}. Moreover, the IDMI-originated nonreciprocity was also predicted for the SW propagation along domain walls in curved stripes \cite{garcia15prl}. Information about the irreversibility of SW properties were also shown in our previous experimental work \cite{Dhi22}.

In this paper, we present the results of experimental and theoretical studies on the irreversibility of magnetization static and dynamic properties in asymmetric Ir/Co/Pt multilayer system with significant IDMI contribution at low uniaxial anisotropy ($Q<1$). Static magnetic properties, characterizing the multilayer, are presented in Section 2. Section 3 shows the magnetic field driven hysteresis of the SW frequency, observed in the BLS experiment. Micromagnetic simulations (enabling determination of the field dependence of magnetization distribution and SW dispersion characteristics) are used, in Section 4, to describe and explain magnetization curves and hysteresis of BLS spectra. Finally, conclusions are presented in Section 5.

\section{Static magnetic properties}

In order to find the magnetic state in the studied [Ir/Co/Pt]$_6$ multilayer (see the description in Methods), firstly, we performed vibration sample magnetometry (VSM) measurements. From the comparison of the magnetization curves measured for the in-plane ($H$) and out-of-plane ($H_\perp$) magnetic fields (see Figure 1(a)) one may deduce an easy plane anisotropy. It is also confirmed by the negative sign of the effective magnetic anisotropy field, $\mu_0\Hueff = -0.4$ T determined from the measurements using an X-band ferromagnetic resonance (FMR) spectrometer. From the superconducting quantum interference device (SQUID) magnetometer, the value of magnetization saturation $\Ms=1.2$ MA/m was determined. Thus the quality factor can be calculated, $Q=0.73$.

\begin{figure*}[tbh!]
\centering
\includegraphics[width=\textwidth]{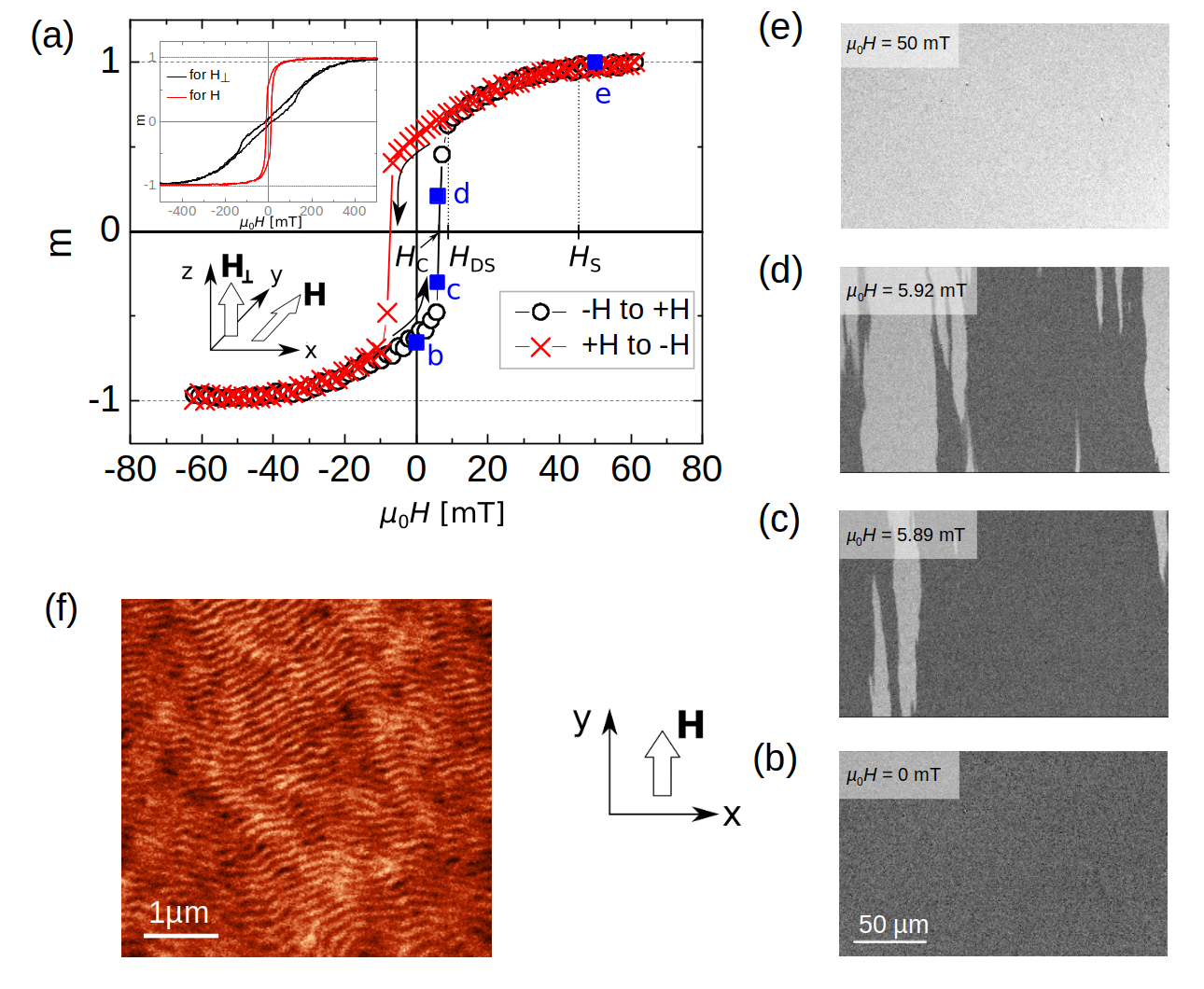}
\caption{Characterization of the sample static magnetization properties: (a) LMOKE hysteresis loop for increasing and decreasing in-plane magnetic field ($H$), the open circles and crosses, respectively. The inset presents VSM hysteresis loops measured at out-of-plane $H_\perp$ and in-plane $H$ magnetic fields; (b--e) a sequence of domain structure images registered by LMOKE (marked with letters and blue squares in the LMOKE curve) illustrating magnetization process starting from in-plane saturated sample with $\mu_0H < 0$; images were recorded at gradually increasing field: (b) $\mu_0H$ = 0 mT; (c) 5.89 mT; (d) 5.93 mT, and (e) 50 mT; (f) domain structure registered by MFM at zero field.}
\end{figure*}

The results of magnetization reversal driven by in-plane magnetic field $H$ studied by longitudinal magneto-optical Kerr effect (LMOKE) are shown in Figure 1. Figure 1(a) shows the hysteresis loop with an abrupt magnetization switching at the coercive field, $\mu_0H_\mathrm{C} = +6.0$ mT. Further increase of $H$ enlarges the in-plane magnetization component up to the saturation field $\mu_0H_\mathrm{S}$, which occurs at about 40 mT. Figures 1(b--e) illustrate the magnetization reversal process while increasing $\mu_0H$ from $-60$ mT ($H$ is applied along the $y$-axis). The reversal process can be described as follows. The sample is initially in a monodomain state with the magnetization component aligned along the $-y$ axis ($M_\mathrm{y} < 0$); see the dark monodomain state in Figure 1(b). While $H$ is approaching $H_\mathrm{C}$, the nucleation of brighter domains (with $M_y > 0$) is observed. Further increase of $H$ causes domain walls movement resulting in the increase of brighter domain area, see Figures 1(c) and (d). The size of these bright and dark domains is of the order of several tens of micrometers. The domains have an anisotropic shape elongated in the direction of $H$. The darker domains vanish at domain saturation field $H_\mathrm{DS}$, slightly higher than $H_\mathrm{C}$. The monodomain state with the magnetization fully oriented along $H$ is achieved at $H_\mathrm{S}$, e.g. Figure 1(e). The magnetization curve is reversible in the field range between $H_\mathrm{DS}$ and $H_\mathrm{S}$. A very different image of the domain structure appears in the studies of MFM, see Figure 1(f). The maze-type domains observed at remanence indicate that: (i) their out-of-plane magnetization components are alternating; (ii) a similar width of the alternating domains means an out-of-plane demagnetized state; (iii) a domain period is of about 100 nm. The following conclusions can be deduced from LMOKE and MFM domain studies: (i) the existence of large macro-domains (with several tens of micrometers in size) with an opposite in-plane magnetization either along $+y$ or $-y$ axis; (ii) the macro-domains modulated in hundred-nm scale with alternating out-of-plane magnetization component aligned along the $\pm z$ axis. Similar stripe in-plane domains periodically modulated with up and down magnetization were also observed in thick enough magnetic films and multilayers without IDMI \cite{maziewski14pssa,fallarino19prb}.

\section{Dynamic magnetic properties}

The influence of $H$ on SW dynamics is investigated using BLS in the backscattering configuration (Figure 2(a)). Here, we choose DE geometry in which SWs propagate perpendicular to $H$ and the IDMI effect on SW is the most pronounced. The measurements are performed in the field range $-400$ < $\mu_0H$ < 400 mT. As it can be seen in Figure 2(a), the investigated SW wavevector lies in the plane of incidence and its length $k = 4\pi\sin\theta/\lambda$, where $\theta$ is an angle between the laser beam of wavelength $\lambda$ and the normal to the sample plane. The rotation of the sample around a planar axis enables the selection of $k$ in the range from 4 to 21 rad/$\mu$m. Figures 2(b) and (c) show the exemplary BLS spectra of SWs, measured for $k$ = 11.8 rad/$\mu$m at $\mu_0H$ equal to (b) 0, and (c) $-58.8$ mT, respectively. To define the positions of Stokes and anti-Stokes peaks the experimental points are fitted by the Lorentzian function (the black lines).

\begin{figure*}[tbh!]
\centering
\includegraphics[width=\textwidth]{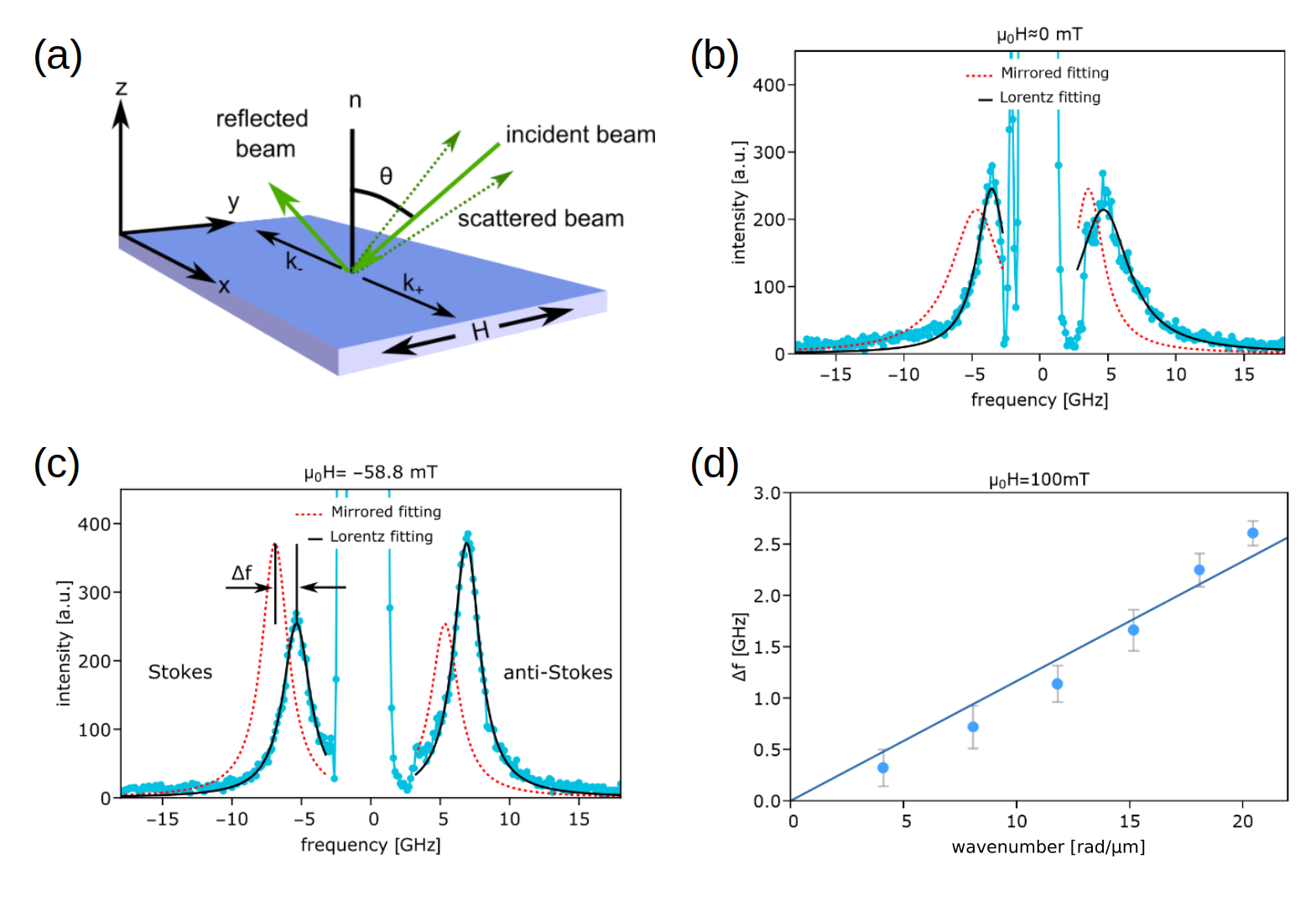}
\caption{BLS measurements. (a) A scheme of backscattering geometry for the spectroscopy of surface SWs propagating in the sample with in-plane magnetization. (b) and (c) Exemplary BLS spectra (Stokes $\fs$ and anti-Stokes peaks $\fas$) measured at the incidence angle $\theta= 30^\circ$ ($k$ = 11.8 rad/$\mu$m) for $\mu_0H$ = 0 mT and $-58.8$ mT, respectively. The black lines represent Lorentzian fits to the experimental points. The red dotted lines (mirror curves to the black ones) enable the intuitively determination of $\Delta f$, the difference between $\fs$ and $\fas$ frequencies. (d) $\Delta f$ dependence on wavenumber $k$ for the sample magnetized by $\mu_0H$ = 100 mT, determined by the BLS measurements (the blue dots) and the linear fit (the solid line), which allows for determination of $\Deff$ using Eq.~(2).}
\end{figure*}

The SW dispersion relation of the DE mode in media with IDMI is described \cite{koerner15prb,di15prl, kostylev14jap, belmeguenai15prb} by the following equation
\begin{equation}
f(k) = f_0(k)+f_\mathrm{IDMI}(k) ,
\end{equation} 							
as a sum of the ``classical'' symmetrical contribution $f_0(k)$, and the asymmetrical $f_\mathrm{IDMI}(k)$ component, 
$f_\mathrm{IDMI}(k)=(\pm\gamma\Deff k)/(\pi \Ms)$, 
where $\gamma$ is the gyromagnetic ratio, $\Deff$ is the effective IDMI constant. The ``$\pm$'' sign is related to the direction of magnetization (along $+y$ or $-y$) and/or direction of wave propagation ($k$ along $+x$ or $-x$, Stokes/anti-Stokes). Equation (1) is valid for a uniformly magnetized ultrathin film, and for our study it is treated as an approximation. According to this, from the above formula one can find the difference between the Stokes $\fs$ and anti-Stokes $\fas$ peaks frequencies
\begin{equation}
\Delta f(k) = \fas(k) - |\fs(k)| =\frac{2\gamma}{\pi\Ms}\Deff k ,
\end{equation}
which is clearly related to the presence of IDMI. The difference between Stokes and anti-Stokes peaks is indicated in Figures 2(b--c), as the spectrum and the mirror curves (the solid black and dotted red lines). 
Using Equation (2) we can make a linear fit to the measured dependence $\Delta f(k)$, as presented in Figure 2(d), and determine the value of $\Deff=1.3$ mJ/m$^2$.

\begin{figure*}[tbh!]
\centering
\includegraphics[width=\textwidth]{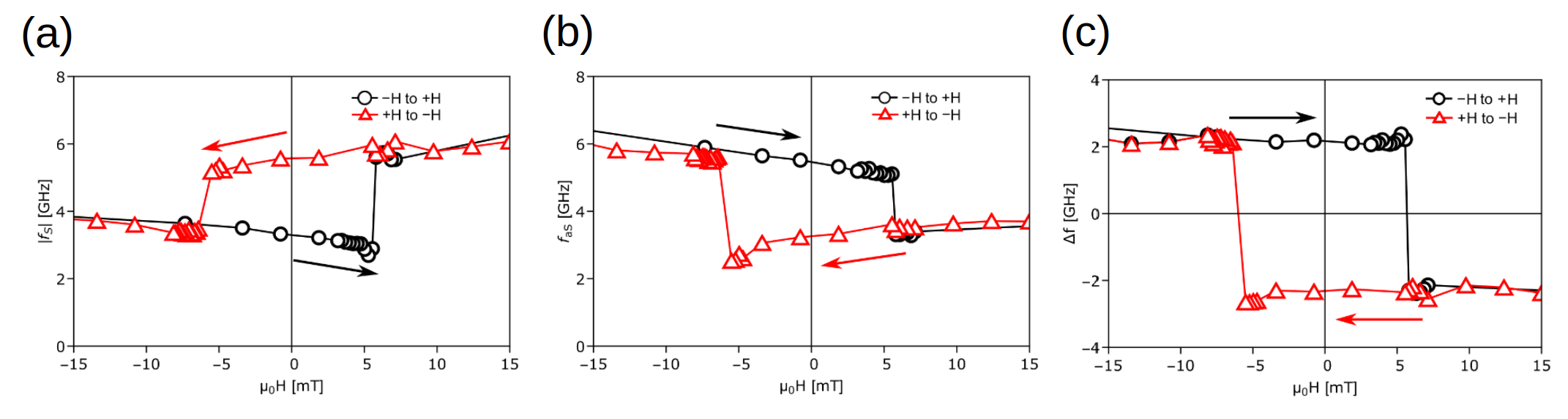}
\caption{$|\fs|$, $\fas$, and $\Delta f$ frequencies as a function of the in-plane magnetic field $H$, i.e., the hysteresis of both Stokes $|\fs|$ (a) and anti-Stokes $\fas$ (b), as well as $\Delta f= \fas- |\fs|$ (c) frequency peaks. The open circles and triangles correspond to the frequencies obtained while increasing and decreasing $H$, respectively. The measurements were performed for $k$=20.5 rad/$\mu$m. }
\end{figure*}

Figure 3 shows the dependences of the Stokes $|\fs|$ (a), anti-Stokes $\fas$ (b), and $\Delta f$  (c) SW frequencies as a function of $H$, in low field regime. It can be observed that as $H$ increases (starting from $-15$ mT), $|\fs|$ and $\fas$ decrease monotonically until the coercivity field $H_\mathrm{C}$. At $H = +H_\mathrm{C}$, a jump in the dependences of $\fas(H)$ and $|\fs|(H)$ occurs, i.e., the frequency $\fas$ decreases and the frequency $|\fs|$ increases. A further increase of $H$ leads to a monotonic increase in $\fas$ and $|\fs|$. For a decrease of the field value starting from $\mu_0H=\mu_0H_\mathrm{C}$ = +15 mT, the process is exactly opposite, and a step-change in the frequency value occurs for $H=-H_\mathrm{C}$. It is worth noting that the step-changes in the $\fas(H)$ and $|\fs|(H)$ are related to the switching of the in-plane magnetization that occurs at $H_\mathrm{C}$, resulting in the change of $\Delta f$ sign.

In the further part of the manuscript we will answer the following questions: what is the relationship between SW dynamics and magnetic configuration in multilayers with IDMI, and how do these relationships depend on the magnetic field. To answer these questions, we will use micromagnetic simulations and confront their results with the experimental measurements, first analyzing how the magnetic configuration changes with the field changes, then analyzing how the dispersion relation changes with it, and finally combining them into a consistent picture.

\section{Discussion}

Micromagnetic simulations are a very powerful tool for the description of magnetic static and dynamic phenomena of complicated systems such as magnetic multilayers. However, to run a simulation, one needs to know a series of material parameters that characterize the system. In our case, the magnetization saturation, the magnetic anisotropy, and magnitude of IDMI were determined experimentally.  
A real value of exchange coupling can be different from the one of bulk material, because the samples deposited by magnetron sputtering on substrates at room temperature are polycrystalline and indicates roughness depending on buffer layer thickness \cite{MatczakJAP13}. Due to such morphology of Ir/Co/Pt multilayers and partial spin polarization of Pt and Ir layers, as well as due to possible interlayer coupling, determining the parameters necessary for micromagnetic simulations is very difficult. Thus we decided to determine them in a combined fitting procedure. We estimate the values of $A$ (exchange coupling) and $S$ (scaling factor $S = A_\perp/A$ between perpendicular inter-layer exchange coupling $A_\perp$ and intra-layer exchange coupling $A$), based on the experimental results of the static LMOKE hysteresis curve (Fig.~1). A best-fit set of the estimated parameters is $A$ = 8.3$\pm$1.0 pJ/m, $S$ = 0.10$\pm$0.02, and all subsequent calculations are presented for these values. 

\begin{figure*}[tbh!]
\centering
\includegraphics[width=\textwidth]{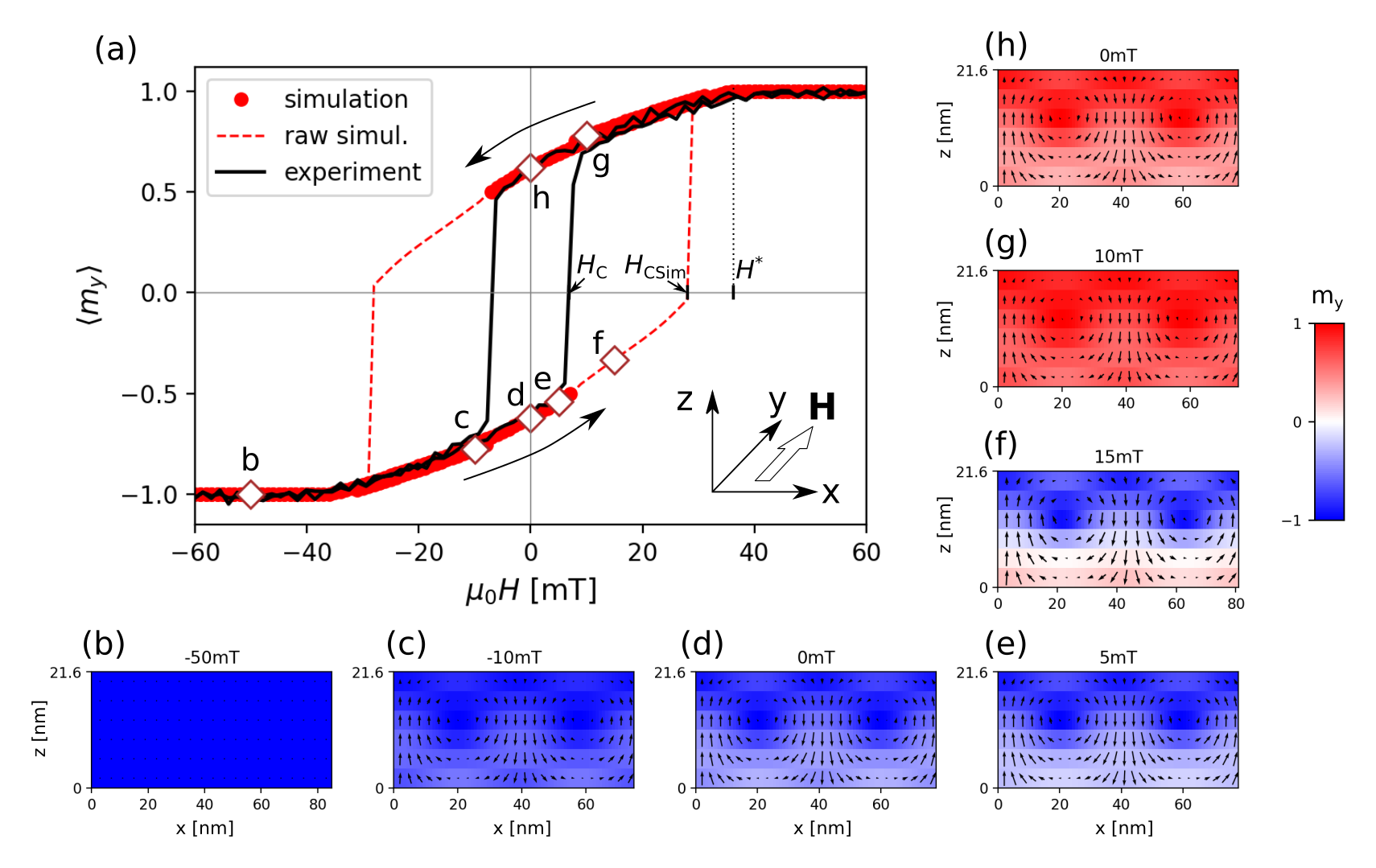}
\caption{The static magnetization properties of the simulated multilayer. (a) The hysteresis loop of $m_y$ component driven by in-plane field $H$ applied along the $y$-direction. The red points correspond to the simulation’s results, whereas the black lines -- to the experimental measurements. For the dashed red curve of large coercivity, see the main text. (b--h) Images show a sequence of the simulated magnetization distributions at the points distinguished in the curve (a) by the enlarged markers. In all the magnetization distribution maps, the magnetization components $m_x$ and $m_z$ are illustrated by arrows whereas the in-plane $m_y$ component is marked by the color scale, see the inset.}
\end{figure*}

Figure 4(a) shows the in-plane applied magnetic field $H$ driven dependence of the mean value of the $y$-component of the reduced magnetization $\langle m_y\rangle = \langle M_y\rangle/\Ms$, reproduced by the micromagnetic simulations, and the experimental LMOKE curve taken from Figure 1(a). These curves are accompanied by the 2D profiles of magnetization distributions for the selected values of $H$. The value of $m_y$ component is coded by the colors defined in the color bar (changes from blue to red correspond to changes of $m_y$ from $-1$ to 1, respectively). $m_x$ and $m_z$ components are described by arrows. 

Figure 4(b) shows the sample fully saturated by negative field (homogeneous blue color, $m_y = -1$). As the field is increased, crossing the critical field, $-H^*$ (close to $-H_\mathrm{S}$, determined from static magnetization curves), $m_x$ and $m_z$ increase and $m_y$ decreases (see Figs. 4(c--e)), and a hybrid domain structure appear, with core magnetization aligned along the $-y$ axis and domains in which the magnetization begins to tilt up or down. This distribution depicts a stripe domain structure oriented along the $y$-axis with a periodicity of 80 nm, close to the maze domain size experimentally determined by MFM imaging, see Figure 1(f). Analyzing the magnetization components $m_x$, $m_z$, one can find vortex-like flux-closure domain walls (with magnetization cores along the $y$-axis) separating domains with $m_z > 0$ and $m_z < 0$. Because of IDMI the magnetization distribution is asymmetric in relation to the central horizontal plane \cite{Legrand18}. As the field is further increased, our simulations indicate that the magnetization switches first in the lower layers (see Figure 4(f)). Finally, at $H_\mathrm{CSim}$ (simulated coercivity field), near $H^*$, the whole sample switches its core magnetization towards $+y$ direction. Note, however, that the blue layers in the upper part of the system shown in Figure 4(f) (in the field well above the experimental coercivity $H_\mathrm{C}$) are in a metastable state, therefore, even a small addition of a local coupling between layers, not considered in our simulations, could lead to whole sample switching and reduction of the coercivity field. So, the difference between $H_\mathrm{C}$ and $H_\mathrm{CSim}$ could be connected with more complicated (than in simplified model used in the simulations) mechanisms of switching observed in the experiment. For instance, defects or local coupling between layers may be responsible for switching of layers in lower fields (with the magnetization oriented in opposite direction to the applied field, blue color in Figures 4(e) and (f)). After saturating the sample in $H > H^*$ and reducing the field afterwards, the process looks similarly: the magnetization distributions presented in Figure 4(g) and (h) are equivalent to the ones in Figures 4(c) and (d), respectively, with opposite $m_y$ component. Particularly, Figures 4(d) and (h) show remnant magnetization distribution characterized by $\langle m_y\rangle = -0.6$ and $\langle m_y\rangle = 0.6$, obtained for the sample initially saturated by large negative and positive field, respectively. 

The calculated dispersion relation $f(k)$ at 100 mT is plotted in Figure 5(a). The intensity of the oscillations of the out-of-plane component of magnetization ($m_z$) of given $(k,f)$ is illustrated with white and green color scale, the white dots are BLS experimental data points as a reference. Since this field is well above the saturation, the magnetization is homogeneously oriented along the $y$ direction, and the simulated spectrum is relatively simple, consisting of two bands that are clearly split for higher $k$. The more intense higher frequency band corresponds to the DE SWs with the amplitude concentrated at one of the surfaces (cf.~mode profiles in Figures 5(c) and (e)), whereas the second band represents the first perpendicular standing SW mode (cf.~mode profiles in Figures 5(d) and (f)). Similarly, Figure 5(b) shows the calculated zero field dispersion relation, which is much more complicated. The presence of the increased number of bands with respect to the homogeneous sample is related to the magnetization texture-induced periodicity and is expected for magnonic crystals \cite{krawczyk14jpcm, banerjee17prb, gruszecki19ssp}. Nonetheless, looking at the intensity of $m_z$ oscillations in $(f,k)$ space and the mode profiles shown in Figures 5(g--j), plotted for SWs at $|k|$=20.5 rad/$\mu$m, it is possible to explain why in the BLS measurements only one band is detected. For the positive (as well as negative) values of $k$, the higher frequency mode is more intense in the simulated dispersion (Figure 5(b)) and its mode profile shows a strong amplitude at the top and bottom surface (see Figure 5(i)), while the lower frequency mode has much smaller amplitude in the simulated dispersion (Figure 5(b)) because the regions with high amplitude of $m_z$ are relatively narrow due to a peculiar distribution of SW amplitude, moreover, $m_x$ is in antiphase at the top and bottom surfaces. 

\begin{figure*}[tbh!]
\centering
\includegraphics[width=\textwidth]{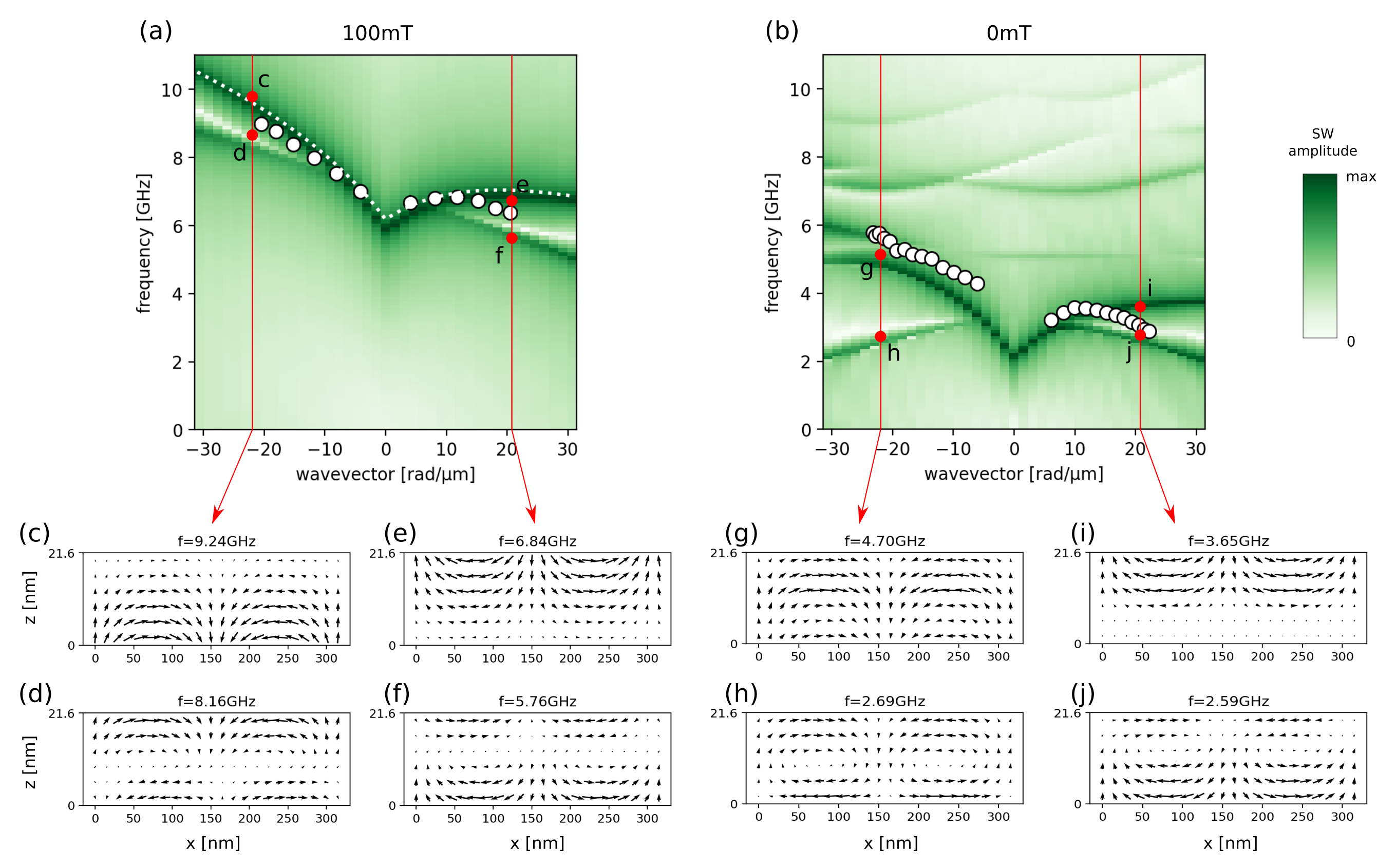}
\caption{The dispersion relations for (a) the uniformly magnetized system with in-plane applied magnetic field of $\mu_0H$ = 100 mT and (b) the system in remanence ($H = 0$) after sample saturation. The white circles represent the experimental data from the BLS measurements. The color maps in the background represent the simulated SW spectra, a color bar at right serves as a scale. With the white dotted line in (a) the theoretical dispersion relation is plotted, according to Eq.~(2). The snapshots of SW profiles corresponding to $k = -20.5$ rad/$\mu$m for the system at $\mu_0H$=100 mT and 0 are displayed in (c--d) and (g--h), respectively, whereas for $k$ = +20.5 rad/$\mu$m for the system at field 100 mT and 0 are shown in (e--f) and (i--j), respectively.}
\end{figure*}

The dispersion relations $f(k)$ were also calculated for a broad magnetic field range. For each field value one can extract the dominating frequency value that corresponds to the experimental value, measured at $k$ = 20.5 rad/$\mu$m. Such series of $f(H)$ values is presented in Figure 6(a), and compared with the analogous experimental dependence (with the pale open points), as presented in Figure 3, yielding a good agreement. A hysteretic behavior is visible, dependent on increasing or decreasing magnetic field. The origin of this mechanism can be found in a series of the calculated dispersion relations for some selected field values (frequency values of one point in each dispersion relation contribute to the presented $f(H)$ dependence). One can see that the dispersions within the upper and lower branches have opposite symmetries (the slopes of bands are opposite), but in each case the selected point comes from the right hand side band, for $k > 0$ ($k$ = 20.5 rad/$\mu$m). Obviously, the bands are ``flipped'' horizontally due to the change of the sign of the static $m_y$ magnetization component, depending on the magnetic history of the sample, as shown in Figure 4. While switching the magnetization, a relation between $k$ and $M$ vectors is changed, so either ``+'' or ``--'' sign from the expression for $f_\mathrm{IDMI}$ is realized. 

\begin{figure*}[tbh!]
\centering
\includegraphics[width=\textwidth]{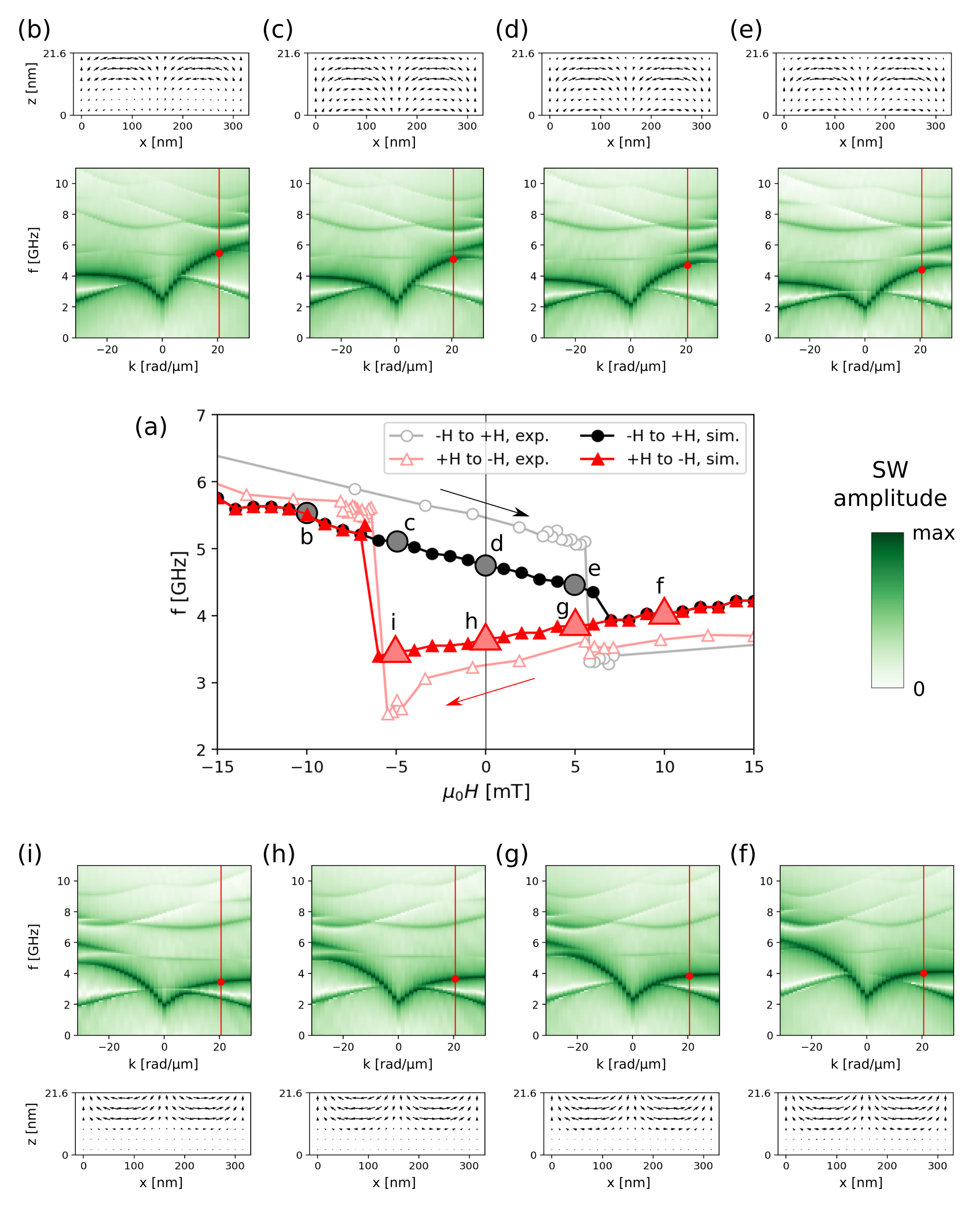}
\caption{The magnetic field-dependent hysteresis of SW frequency, calculated using the micromagnetic simulations. (a) $f(H)$ dependence, constructed from the frequency values for $k$ = 20.5 rad/$\mu$m from the highest intensity band in the dispersion relations. The dispersion relations for some selected field values are plotted at the top (b--e) and the bottom (f--i) rows, and resulting frequencies are indicated in (a) with enlarged markers. Colors in (b--i) are normalized in the same way, therefore, the values corresponding to these colors (cf. colorbars) are proportional to magnitudes of SWs at given $H$ and $f$. The experimental frequency values from BLS are displayed with the pale open points. Each dispersion relation in (b--i) is assisted with a profile of the oscillation modes for $k$ = 20.5 rad/$\mu$m, from the most intensive band.}
\end{figure*}

The snapshots of the dynamic magnetization profiles corresponding to SWs of selected frequency and wavenumber are presented at the top of Figure 6 for four values of $H$ as it increases from $-H$ to $+H$  (b--e) and at the bottom, as it sweeps from $+H$ to $-H$ (f--i). It is clearly seen that these mode profiles have a typical character of DE modes with the amplitude concentrated at top or bottom interfaces (different for waves propagating in the opposite directions). It is also evident that SWs propagating in the opposite directions have different chirality what explains the IDMI-induced frequency shift between them.

\begin{figure*}[tbh!]
\centering
\includegraphics[width=\textwidth]{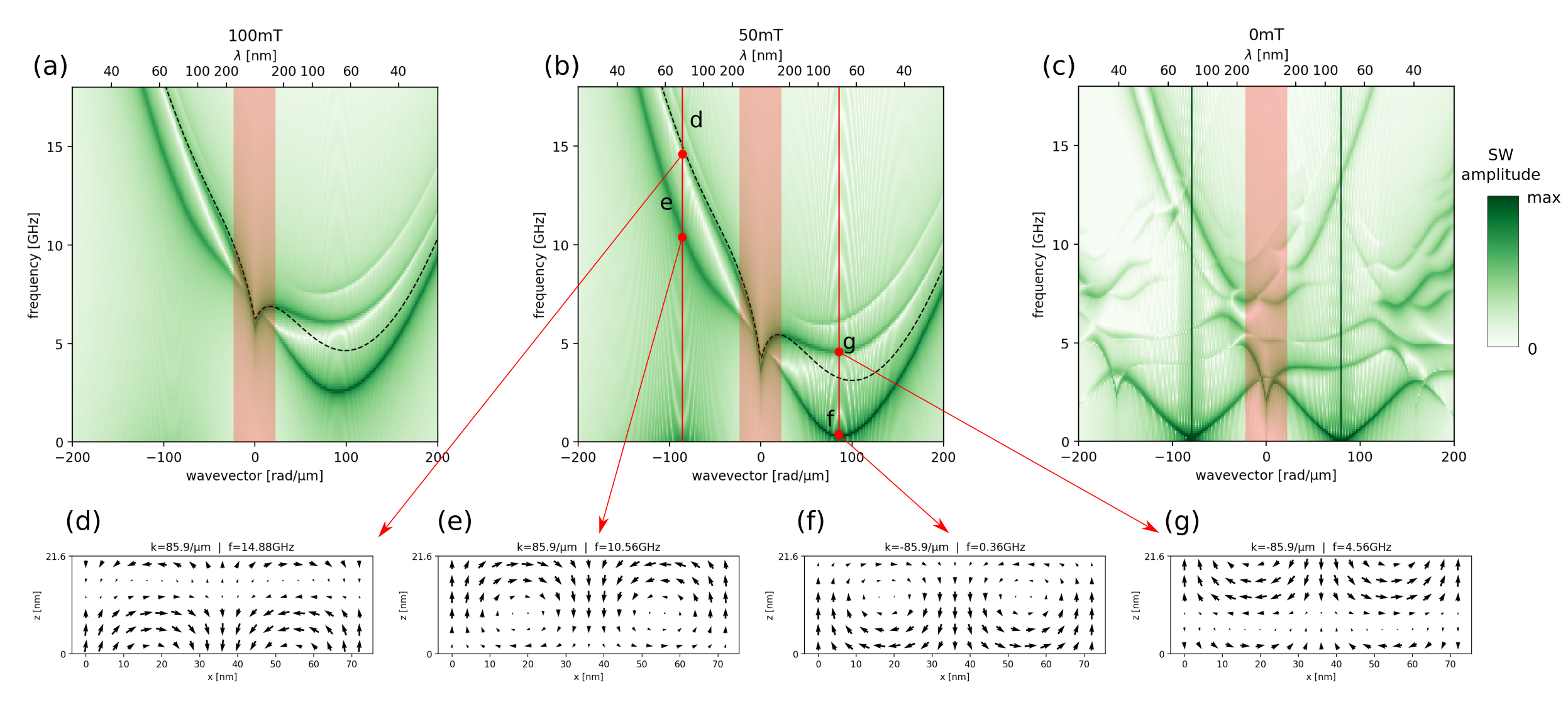}
\caption{Dispersion relations computed for a broad $k$ range, for: (a) and (b) uniformly magnetized structure by the field values of 100 mT and 50 mT, respectively and (c) stripe domain pattern at the remanence state. Colors in (a--c) are normalized in the same way, therefore, the values corresponding to these colors (cf.~colorbars) are proportional to magnitudes of SWs at given $k$ and $f$. (d--e) and (f--g) depict the mode profiles for SW in the uniformly magnetized system at $\mu_0H=50$ mT, for $k$ = -85.9 rad/$\mu$m and $k$ = 85.9 rad/$\mu$m, respectively. Precise values of $k$ and $f$ of these modes are listed above these plots. The shaded area in the region of small $k$ values corresponds to the accessible range of BLS measurements. With the black dashed lines in (a) and (b) the theoretical dispersion relations, according to Eq.~(1), are plotted for comparison.}
\end{figure*}

Figure 7 shows the simulated dispersion relations for a wide range of $k$ at three values of $\mu_0H$: 100, 50 and 0 mT, taking into account the IDMI contribution. Strong changes in the dispersion characteristics are observed between monodomain and multidomain states, which appear below the saturation field $H^*$ (compare Figures 7(b) and (c)). Figures 7(a, b) are for a monodomain state and illustrate the IDMI-induced asymmetry in the dispersion relation resulting in differences in SW velocities for opposite propagation directions. Zero SW group velocity is obtained at $H^*$, when $k$ in the dispersion minima approaches the critical value $k^*$ (SW mode softening), as shown in Figure 7(b) (for the field slightly above $H^*$). Here, due to IDMI, it appears only for one propagation direction, which is positive in the discussed case. For this direction the phase velocity is strongly changeable, and it decreases also to zero when SW is frozen into domain state \cite{Kis23PRB}. While changing the applied field direction, the dispersion characteristic undergoes mirror like transformation. Figures 7(d--g) show mode profiles at characteristic points of this dispersion relation. Particularly interesting case is SW mode softening (Figure 7(f)), as it resembles the static vortex-like flux-closure magnetic configuration, presented in Figure 4. The mode profile for opposite value of $k = -k^*$ (Figure 7(e)) has the same shape, flipped vertically, but here the frequency in much higher. Figure 7(c) shows the dispersion relation at remanence with visible periodicity caused by the periodicity of the magnetization texture, confirming that our system behaves like a magnonic crystal at remanence, however, without full band gaps in the considered frequency range. 

\section{Conclusions}

The influence of IDMI on magnetization statics and dynamics in [Ir/Co/Pt]$_6$ multilayer with $Q < 1$ were studied experimentally and using micromagnetic simulations, in the transition from uniform magnetization to stripe domain pattern. The following static stripe domain magnetization distribution was deduced: large macro-domains (several micrometer size) with in-plane ``core'' magnetization which are divided into small nano-domains (about 100 nm size) differentiated by a sign of the out-of-plane magnetization component. The hystereses of the Stokes, anti-Stokes peaks, and $\Delta f$ frequencies as a function of the in-plane applied magnetic field were found from BLS studies. The hystereses are explained combining the IDMI effect and macro-domains switching. The IDMI-induced large differences in group and phase velocities of SWs, propagating in opposite directions, are deduced from the SW dispersion relations measured both in a field-created monodomain state as well as in a remnant state. Overall, we have demonstrated that the multilayers with IDMI and low uniaxial anisotropy have very interesting properties and can be treated as a nonreciprocal, reprogrammable medium for SW propagation, which can be crucial for applications in magnonics and spintronics, for example serving as tunable nonreciprocal magnonic crystals at remanence. Such system could also be used for such applications as, e.g.~low energy switchers based on the hysteresis of Stokes, anti-Stokes frequency peaks.

\section*{acknowledgement}

This work is supported by Polish National Science Center project M-ERA.NET 3 No.~2022/04/Y/ST5/00164, and project 2019/35/D/ST3/03729. The computations were carried out at the Computer Center of University of Bialystok.

\section*{Methods}

\subsection*{Films preparation}

Ferromagnetic multilayers were deposited by DC magnetron sputtering on naturally oxidized Si substrates covered with Ta (4nm)/Au(30nm) buffer layers of nominal structures: [Ir(1 nm)/Co 1.6 nm)/Pt(1 nm)]$_6$. The deposition was done at room temperature in a multichamber system with base pressure below 2 $\times$ 10$^{-8}$ mbar and an Ar atmosphere of pressure of 1.4 $\times$ 10$^{-3}$ mbar.

\subsection*{Experimental methods}

In the BLS spectroscopy, a single-mode green laser of wavelength $\lambda$=532 nm was used for measurement of thermal excitation of SWs. The Stokes peak -- a creation of magnon (negative frequency mode $\fs$) and anti-Stokes peak -- an annihilation of magnon (positive frequency mode $\fas$), in photon-magnon interactions were measured. The scattered light was analyzed with a multi pass Tandem Fabry-Perot TFP-2 HC Interferometer. Our results are supported by complementary measurement techniques such as: (i) longitudinal magneto-optical Kerr effect (LMOKE) based microscopy sensitive to the in-plane magnetization component; (ii) magnetic force microscopy (MFM); (iii) X-band ferromagnetic resonance FMR spectrometer.

\subsection*{Micromagnetic simulations}

Micromagnetic simulations were performed using Mumax3 software \cite{vansteenkiste14aip}. To model the multilayer system with a magnetic film and a nonmagnetic spacer we exploited a solution proposed by Woo et al. \cite{woo16nat, Joos23}, with effective magnetic parameters $\Ms'=\varepsilon \Ms$, $\Aex'=\varepsilon \Aex$, $\Deff' =\varepsilon \Deff$, and $K_\mathrm{eff}'=\varepsilon K_\mathrm{eff}$ (effective anisotropy constant, $K_\mathrm{eff}=K_\mathrm{u}-\frac{1}{2}\mu_0\Ms^2$), where $\varepsilon$ is the a fraction of the magnetic film volume in the multilayer system, here $\varepsilon$ = 1.6 nm/(1.6 nm+2.0 nm)=0.44. It should be noted, that all described magnetic parameters are referred in the manuscript as for pure magnetic film (without apostrophes), but the scaled values for the diluted medium (with apostrophes) are used to simulate an effective medium. The system of total thickness (along the $z$-axis) equal to (1.6 nm+2 nm)$\times$6 = 21.6 nm was discretized by the unit cells of dimensions $c_x \times c_y \times c_z = 1 \times 1 \times 3.6$ nm$^3$. In order to consider our system as infinite in the $(x,y)$-plane, we have assumed the periodic boundary conditions along the $x$ and $y$-axes. A different value of the exchange interaction, $A$, for neighboring magnetic moments laying in the same plane (the cells with the same $z$-coordinate) and perpendicular to it, $A_\perp$ (the cells with different $z$ coordinate) were assumed, $A_\perp= S\cdot A$. The experimental values of magnetization saturation and effective magnetic anisotropy were used, $\Ms$ = 1.2 MA/m, $\mu_0\Hueff$ = $-0.4$ T, thus the quality factor $Q = 0.73$. The value of $\Deff=1.3$ mJ/m$^2$ was determined from the BLS studies. The other magnetic quantities, namely $A$ and $S$, where determined in a fitting procedure: the hysteresis curve was simulated for combinations of $A$ and $S$ values, and the result was compared with the LMOKE result. The best-fitted parameters were estimated as $A=8.3\pm 1.0 $pJ/m and $S= 0.10\pm0.02$.  
The preferred static magnetization distribution was determined using energy density minimization. The system size was $N_x\times N_y \times N_z = 2048\times 1\times 6$ cells, so with the periodic boundary conditions along $x$ and $y$ directions, the magnetic configuration was infinite stripe domain structure, oriented along the $y$ direction. An initial magnetization distribution was prepared, imposing periodicity of $N_\mathrm{d}$ domains within the constant distance of $N_x$ cells. Then the system was relaxed, and the total system energy density $E$ was calculated, as a function of $N_\mathrm{d}$. The minimal energy $N_\mathrm{d,min}$ value was determined, and the domain period $p = N_x/N_\mathrm{d,min}$ was estimated. In order to calculate the hysteresis curve, these steps were repeated for a series of magnetic field values, applied along the $y$ direction, what corresponds to the LMOKE measurements. For each field value a $m_y$ magnetization component, averaged over the whole system, was taken into account. To include the magnetic history, which is important in our study, the initial magnetization distribution for each subsequent field step was taken from the previous field step, scaled and repeated within constant $N_x$, in order to obtain the proper $N_\mathrm{d}$ value. Twice bigger system size, with $N_x$ = 4096, was taken in dynamic simulations, and initial magnetization distributions at given field were taken according to the outcome of the static simulations. The magnetization dynamics was triggered by applying the out-of-plane polarized microwave field in the form of: $\mu_0h_z = b_0\mathrm{sinc} (k_\mathrm{cut}x)\mathrm{sinc} (2\pi f_\mathrm{cut} (t - t_0))$, where $b_0$ = 1 mT, $k_\mathrm{cut} = 35$ rad/$\mu$m or $k_\mathrm{cut} = 200$ rad/$\mu$m is the cut-off wavenumber (see the $k$-range in Figures 5--7), 
$f_\mathrm{cut}$ = 25 GHz is the cut-off frequency, and $t_0 = 10/f_\mathrm{cut}$. After integrating the time-dependent magnetization distribution over the $z$-th coordinate, the $(x,t)$-dependence of $m_z$ was transformed with two-dimensional FFT into the $(k,f)$-space to obtain the dispersion relation. For a given $k$ value, a cross-section of the dispersion relation was taken, and such spectrum was analyzed to determine the frequency $f_\mathrm{max}$ at which the maximal SW amplitude occurs. Such $f_\mathrm{max}$ could be further compared with the value measured by BLS. The mode profiles, as SW snapshots, were calculated by performing the inverse 2D FFT on the dispersion relation (calculated for each of 6 magnetic layers individually, without integrating over the film thickness), multiplied by a binary mask, which selects desired $k_0$ and $f_0$ values. In all dynamic simulations, a Gilbert damping $\alpha$ = 0.01 was used, which is less than the real value, but allows to obtain sharper bands and better spectral selectivity of the result.


%

\end{document}